\documentclass[10pt,a4paper,twocolumn,english,aps,prl,showpacs,floatfix,superscriptaddress,10pt]{revtex4-1}
\usepackage[T1]{fontenc}
\usepackage[latin9]{inputenc}
\usepackage{color}
\usepackage{babel}
\usepackage{geometry}
\usepackage{float}
\usepackage{amssymb}
\usepackage{amsmath}
\usepackage{graphicx}
\usepackage{breakurl}
\usepackage{footnote}
\usepackage[unicode=true,
 bookmarks=true,bookmarksnumbered=false,bookmarksopen=false,
 breaklinks=false,pdfborder={0 0 0},backref=false,colorlinks=true]
 {hyperref}
\hypersetup{pdftitle={Super Spin Hall Angles of Gold Based Alloys},
pdfauthor={Piotr Laczkowski}}

\makeatletter

\begin{document}

\title{Large enhancement of the spin Hall effect in Au by scattering with side-jump on Ta impurities.} 

\author{P. Laczkowski}
\affiliation{Universit\'e Grenoble Alpes, Spintec, CEA-INAC and CNRS, F-38000 Grenoble, France}
\affiliation{Unit\'e Mixte de Physique CNRS-Thales, Univ. Paris-sud, Universit\'e Paris-Saclay 11, 91767 Palaiseau, France}

\author{Y. Fu}
\affiliation{Universit\'e Grenoble Alpes, Spintec, CEA-INAC and CNRS, F-38000 Grenoble, France}

\author{H. Yang}
\affiliation{Universit\'e Grenoble Alpes, Spintec, CEA-INAC and CNRS, F-38000 Grenoble, France}

\author{J.-C. Rojas-S\'anchez }
\affiliation{Institut Jean Lamour, CNRS-Universit\'e de Lorraine, 54506 Vandoeuvre l\`es Nancy, France}

\author{P. Noel}
\affiliation{Universit\'e Grenoble Alpes, Spintec, CEA-INAC and CNRS, F-38000 Grenoble, France}

\author{V.T. Pham}
\affiliation{Universit\'e Grenoble Alpes, Spintec, CEA-INAC and CNRS, F-38000 Grenoble, France}

\author{G. Zahnd}
\affiliation{Universit\'e Grenoble Alpes, Spintec, CEA-INAC and CNRS, F-38000 Grenoble, France}

\author{C. Deranlot}
\affiliation{Unit\'e Mixte de Physique CNRS-Thales, Univ. Paris-sud, Universit\'e Paris-Saclay 11, 91767 Palaiseau, France}

\author{S. Collin}
\affiliation{Unit\'e Mixte de Physique CNRS-Thales, Univ. Paris-sud, Universit\'e Paris-Saclay 11, 91767 Palaiseau, France}

\author{C. Bouard}
\affiliation{Universit\'e Grenoble Alpes, Spintec, CEA-INAC and CNRS, F-38000 Grenoble, France}

\author{P. Warin}
\affiliation{Universit\'e Grenoble Alpes, Spintec, CEA-INAC and CNRS, F-38000 Grenoble, France}

\author{V. Maurel}
\affiliation{Universit\'e Grenoble Alpes, Symmes, CEA-INAC, F-38000 Grenoble, France}

\author{M. Chshiev}
\affiliation{Universit\'e Grenoble Alpes, Spintec, CEA-INAC and CNRS, F-38000 Grenoble, France}

\author{A. Marty}
\affiliation{Universit\'e Grenoble Alpes, Spintec, CEA-INAC and CNRS, F-38000 Grenoble, France}

\author{J.-P. Attan\'e}
\affiliation{Universit\'e Grenoble Alpes, Spintec, CEA-INAC and CNRS, F-38000 Grenoble, France}

\author{A. Fert}
\affiliation{Unit\'e Mixte de Physique CNRS-Thales, Univ. Paris-sud, Universit\'e Paris-Saclay 11, 91767 Palaiseau, France}

\author{H. Jaffr\`es}
\affiliation{Unit\'e Mixte de Physique CNRS-Thales, Univ. Paris-sud, Universit\'e Paris-Saclay 11, 91767 Palaiseau, France}

\author{L. Vila}
\affiliation{Universit\'e Grenoble Alpes, Spintec, CEA-INAC and CNRS, F-38000 Grenoble, France}

\author{J.-M. George}
\affiliation{Unit\'e Mixte de Physique CNRS-Thales, Univ. Paris-sud, Universit\'e Paris-Saclay 11, 91767 Palaiseau, France}

\date{\today}

\begin{abstract}
We present measurements of the Spin Hall Effect (SHE) in AuW and AuTa alloys for a large range of W or Ta concentrations by combining experiments on lateral spin valves and Ferromagnetic-Resonance/spin pumping technique. The main result is the identification of a large enhancement of the Spin Hall Angle (SHA) by the side-jump mechanism on Ta impurities, with a SHA as high as + 0.5 (i.e $50\%$) for about 10\% of Ta. In contrast the SHA in AuW does not exceed + 0.15 and can be explained by intrinsic SHE of the alloy without significant extrinsic contribution from skew or side-jump scattering by W impurities. The AuTa alloys, as they combine a very large SHA with a moderate resistivity (smaller than $85\,\mu\Omega.cm$), are promising for spintronic devices exploiting the SHE.
     
\end{abstract}

\maketitle

A goal of spintronics is to generate, manipulate and detect spin currents for the transfer and manipulation of information, thus allowing faster and low-energy consuming operations. Since the discovery of the Giant Magnetoresistance a "classical" way to produce spin currents was to take advantage of ferromagnetic materials and their two different spin channel conductivities \cite{Baibich:1988he, Grunderg1989}. 
In the last decade the rediscovery and study of spin orbit interaction effects brought new insights into creation mechanisms of spin currents. Among all mechanisms the spin Hall effect (SHE) focused a lot of attention as it allows the generation of spin currents from charge current and $vice\,versa$ \cite{DyakonovPerel1971}. 
Despite being observed only a decade ago \cite{Kato2004, wunderlich2005experimental, Valenzuela} these effects are already ubiquitous within the Spintronics as standard spin-current generators and detectors \cite{Hoffmann2013, NiimiRepProgr2015, Sinova:2015ic}. 
The conversion coefficient between charge and spin currents is called the spin Hall angle (SHA) and is defined as $\Theta_{SHE}= {\rho}_{xy}/{\rho}_{xx}$, ratio of the non-diagonal and diagonal terms of the resistivity tensor. 
One of the main interests of the SHE is to provide a new paradigm for Spintronics where non-magnetic materials becomes active spin current source and detector. 

Until now most of the reports focused on single heavy metals and intrinsic SHE mechanisms, the main materials of interest being: Pt, Ta, W, and some oxydes. With intrinsic mechanisms the SHA is typically proportional to the resistivity of the heavy metal, and generally, a large value of the SHA is associated with a high resistivity (i.e. - 0.3 for the SHA in ${\beta}-W$ is associated with $263\ {\mu}{\Omega.}cm$ \cite{Pai:2012ef}) which limits the current density and the resulting spin transfer torques on the magnetisation of an adjacent metallic ferromagnetic material. Extrinsic SHE mechanisms associated with the spin dependent scattering on impurities or defects are an alternative to generate transverse spin currents \cite{Fert:2011bl}. Two particular scattering mechanisms have been identified: the skew scattering  \cite{Smit1958} providing a non-diagonal term of the resistivity tensor proportional to the longitudinal resistivity (${\rho}_{xy}\propto{\rho}_{xx}$) and the side jump \cite{Berger1970} for which the non-diagonal term is proportional to the square of the resistivity (${\rho}_{xy}\propto{\rho}_{xx}^2$).  For instance, the skew scattering mechanism have been observed in CuIr, CuBi, CuPb alloys (SHA=0.02, -0.24, -0.13, resp.) \cite{Niimi:2011ii, Niimi:2014}. The intrinsic mechanism from Berry curvature in the conduction band gives the same dependence ${\rho}_{xy}\propto{\rho}_{xx}^2$ as the side-jump contribution so that, for example the SHE of AuPt (SHA=0.3 at max.) alloys could be explained by a predominant intrinsic effect rather than ascribed to side-jump\cite{Obstbaum}. 

In this letter we present a study of Au-based alloys with $W$ and $Ta$ impurities. We demonstrate that the side-jump scattering mechanism dominates in AuTa alloys, and generates high spin Hall angles (up to + 0.5) with the additional advantage of resistivities ($\rho _{AuTa}< 85 \mu \Omega.cm$) smaller than in most materials with SHA in the same range. By contrast in AuW alloys the SHE is mainly due to only the intrinsic mechanism and is definitely smaller than in AuTa. This difference between AuTa and AuW is supported by $ab-initio$ calculations.

The alloys were fabricated by DC magnetron sputtering by co-deposition of the two pure materials. The concentration in atomic purcent were determined by chemical analyzes (proton or electron induced X-ray emission) and from the deposition rate of each species. We control the alloying through the increase of the resistivity as the concentration is increased. We found an almost linear relationship, see \cite{Suppl}, as expected for diluted alloys. We note that care has to be taken with heat treatment to avoid eventual clustering of the impurities, that can be detected trough the eventual drop of the nominal resistivity. Experimentally, the spin Hall angle has been characterized by the Inverse Spin Hall Effect (ISHE) using both lateral spin valves with inserted SHE materials and spin pumping ferromagnetic-resonance experiments (SP-FMR) \cite{Tserkovnyak}. For both types of experiments we follow excatly the same experimental protocols described in our previous work \cite{Laczkowski_APL_2014}. For the determination of the spin diffusion length (${\lambda}_{sf}$) by spin absorption experiments in LSV we use an extended model allowing its precise estimation, as detailed in \cite{Laczkowski:2015ge}. 

\begin{figure}
	\centering{}
		\includegraphics[width=7.8cm]{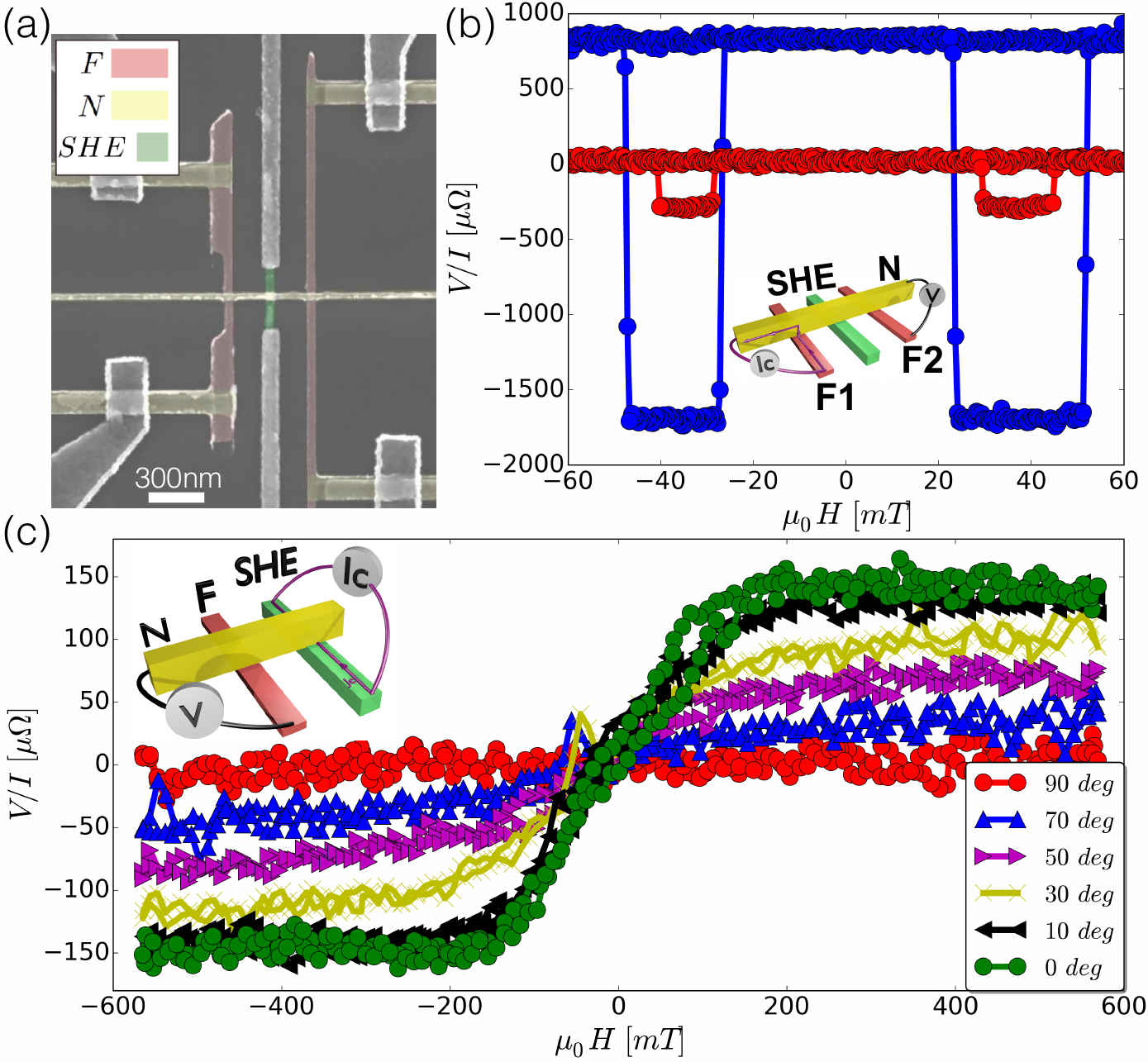}
		\caption{\textit{(a) SEM image of a Non-Local Lateral Spin Valve structure with inserted AuW nano-wire (b) Typical non-local spin signals in experiments of spin absorption by the SHE material(AuW): reference device without AuW (blue) and device with AuW (red). (c) Field dependence of the signal induced by Inverse Spin Hall Effect at different orientations of the field. The insets in (b) and (c) display the respective non local probe configurations.}}
\end{figure}

Figure 1(a) displays a SEM image of a typical lateral spin valve used in the experiments. Schematic representation of the non-local probe configuration is represented in the insets of Fig.1 (b-c). Non-local spin signals recorded for the reference LSV without AuW (blue) and the LSV with spin absorption by the inserted AuW (red) are displayed in the figure 1(b). These are the typical measurements which are used for spin sink experiments and allowing the extraction of the spin diffusion length in the SHE material \cite{Laczkowski:2015ge}. Figure 1(c) represents the angular variation of the Inverse Spin Hall Effect voltage signal as a function of the external magnetic field for a typical device (the voltage is measured between both sides of the inserted AuW nano-wire \cite{vila}). With these measurements one has access to the spin diffusion length of a given AuW alloy and its spin Hall angle. The accuracy of SHE measurements with LSVs is however limited by current shunting effects \cite{Niimi:2011ii,Laczkowski_APL_2014} for alloys of large resistivity, typically for $\rho>100\,{\mu}\Omega.cm$. We have thus used finite element method simulations for the correction of the shunting effect by the non magnetic Cu channel \cite{Laczkowski_APL_2014}.

We also performed measurements of ISHE by SP-FMR \cite{MosendzPRB2010, Azevedo2011, Nakayama2012} at room temperature on $SiO_2\textbackslash \textbackslash Py \textbackslash AuX$ bilayers (X being W or Ta) in a split cylinder microwave resonant cavity. At the ferromagnetic resonance, a pure $dc$ spin current $j_s$ is injected into the Au-based alloy layer along $z$ direction with the spin polarisation along $x$ direction [see Fig. 3(c)]. Due to the ISHE in Au-based alloy this spin current is then converted into a transversal $dc$ charge current along the $y$ direction, or into a transverse $dc$ voltage in an open circuit measurement. The $rf$ magnetic field is applied along the long samples axis and the external applied $dc$ magnetic field $H_{dc}$ along the width [Fig.3(c)]. The frequency of $h_{rf}$ is fixed to $9.75 \, GHz$ whereas $H_{dc}$ is swept around the FMR condition. The amplitude of the $h_{rf}$ was determined for each measurement by measurement of the resonant cavity $Q$ factor with the sample placed inside. The derivative of FMR absorption spectra is measured at the same time as the voltage taken across the long extremity of the sample [Fig.3(a) and (b)]. We have also carried out a frequency dependence ($3-24\,GHz$) of the FMR spectrum in order to determine the effective saturation magnetisation $M_{eff}=760 \pm 30$ kA/m as well as the damping constant ${\alpha}_{NiFe}=6.9 \pm 0.1 10^{-3}$ and the inhomogeneous broadening, see \cite{Suppl}. This allows us to estimate the effective spin mixing conductance $g_{eff}$ ($6 \pm 1 nm^{-2}$ and $4.5 \pm 0.5 nm^{-2}$ for AuW and AuTa, respectively) and thus the spin current density $j_s$ injected by SP-FMR as described in \cite{Ando2011}. 

\begin{figure}
	\centering{}
		\includegraphics[width=7.8cm]{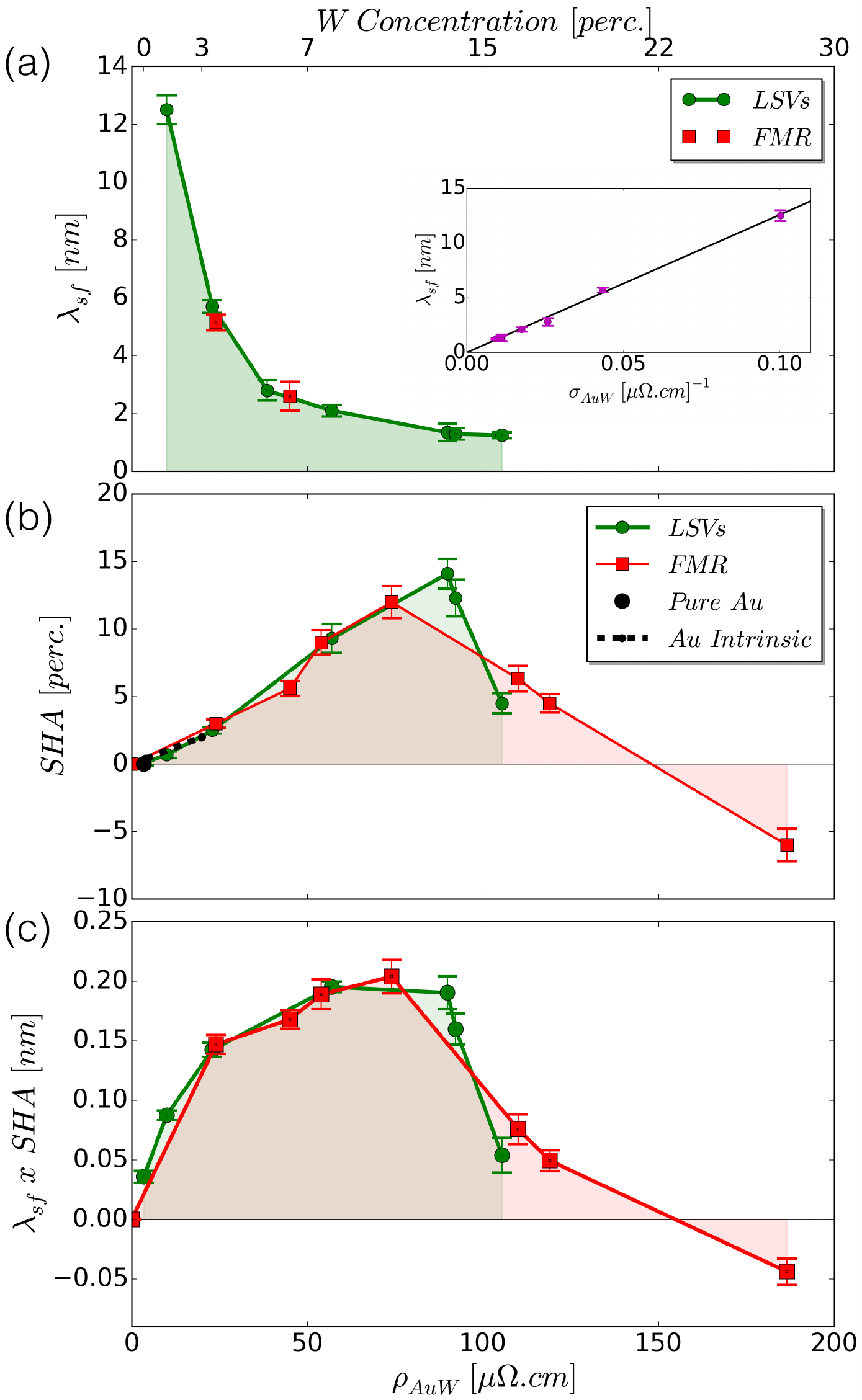}
		\caption{\textit{(a) Spin diffusion lengths extracted using spin-absorption experiments in LSVs (green) or spin pumping voltage in FMR (red). Inset: the $\rho \times \lambda_{sf}$ product remains approximately constant in the whole experimental range. (b) Dependence of the spin Hall angles derived by LSVs  and SP-FMR  techniques on the AuW concentration/resistivity. The dashed line represents the expected intrinsic contribution derived by averaging data on Au films in the $4-19 \mu \Omega.cm$ resistivity range, see text. (c) Same as in (b) for the product of the spin Hall angle and the spin diffusion length characterizing the efficiency of the spin-charge conversion \cite{Suppl,Sanchez:kb}.}}
\end{figure}


We first focus on our results on SHE in AuW alloys. Figure 2(a) displays the spin diffusion lengths evaluated using spin absorption experiments following the protocol described in \cite{Laczkowski:2015ge}. As expected the spin diffusion length decreases as the resistivity of the AuW increases, going from $13\,nm$ at low resistivity to $1.2\,nm$ at higher resistivities. One checks that the $\rho \times \lambda_{sf}$ product remains constant in the whole explored resistivity range [Inset of Fig. 2(a)], as expected if both the spin and momentum scattering rates $1/\tau_{sf}$ and $1/\tau$ increase in the same way with the impurity concentration ($\rho \lambda_{sf} \propto \sqrt{\tau_{sf}/\tau} $). We have also confirmed the spin diffusion lengths extracted from LSVs by using SP-FMR experiments. For this purpose we have studied the evolution of the measured $I_{c}$ as a function of AuW thicknesses, for low resistivity samples and when the spin diffusion length is large enough to observe the variation of the current until its saturation at $2 {\lambda}_{sf}$. The resulted spin diffusion lengths are in perfect agreement with the ones from LSVs, as represented in Fig.2 (a) by red squares. 

Figure 2(b) summarizes the resistivity dependence of the SHA in the AuW alloys derived from LSVs (green) and SP-FMR (red). We can see an almost linear initial increase of the SHA up to about 15\% followed by a decrease when the AuW resistivity reaches $90\ \mu\Omega.cm$ for 13\% of W. The intrinsic SHE mechanism related to the Berry curvature of the conduction band (independently of extrinsic effects from skew or side-jump scatterings) is expected to give such a linear variation, at least in the limit of small concentration of W and small changes of the conduction band. Actually the dashed line in Fig.2(b) represents the intrinsic SHE expected from an average on data we got on pure gold films or have found in the review on SHE of Hoffmann for films of similar thickness (\cite{Hoffmann2013}). This line (slope of 0.1\%/$\mu\Omega.cm$) is close to the experimental variation at small concentration for both the data from LSV and spin pumping. This indicates that the intrinsic SHE is likely the predominant mechanism of SHE in AuW. The change of sign of the SHA at concentrations larger than about 13\% is consistent with a change of sign of the intrinsic SHE between a positive sign for pure Au and a negative one for pure W. 

For concentrations above 13\%, since the SHA is not derived with a great accuracy by LSV experiments in the range where the spin diffusion length becomes small, we have plotted only the SHA derived from spin pumping in Fig.2b. The variation of the length ${\lambda}=SHA \times \lambda_{sf}$ characterizing the yield of spin-charge conversion \cite{Suppl,Sanchez:kb,RojasSanchezPRL216} is shown in Fig. 2c and reaches 0.2 nm.  

After having established that both lateral spin-valves and FMR-ISHE techniques lead to the same spin Hall Angles and spin diffusion lengths, we now focus on the results obtained on AuTa alloys by using only spin-pumping/ferromagnetic resonance experiments. Typical FMR spectra (field derivative of FMR absorption $dX^{''}/dH$) and the corresponding field dependence of the voltage, detected simultaneously, are shown in Fig. 3(a) and 3(b) respectively.

\begin{figure}
	\centering{}
		\includegraphics[width=7.8cm]{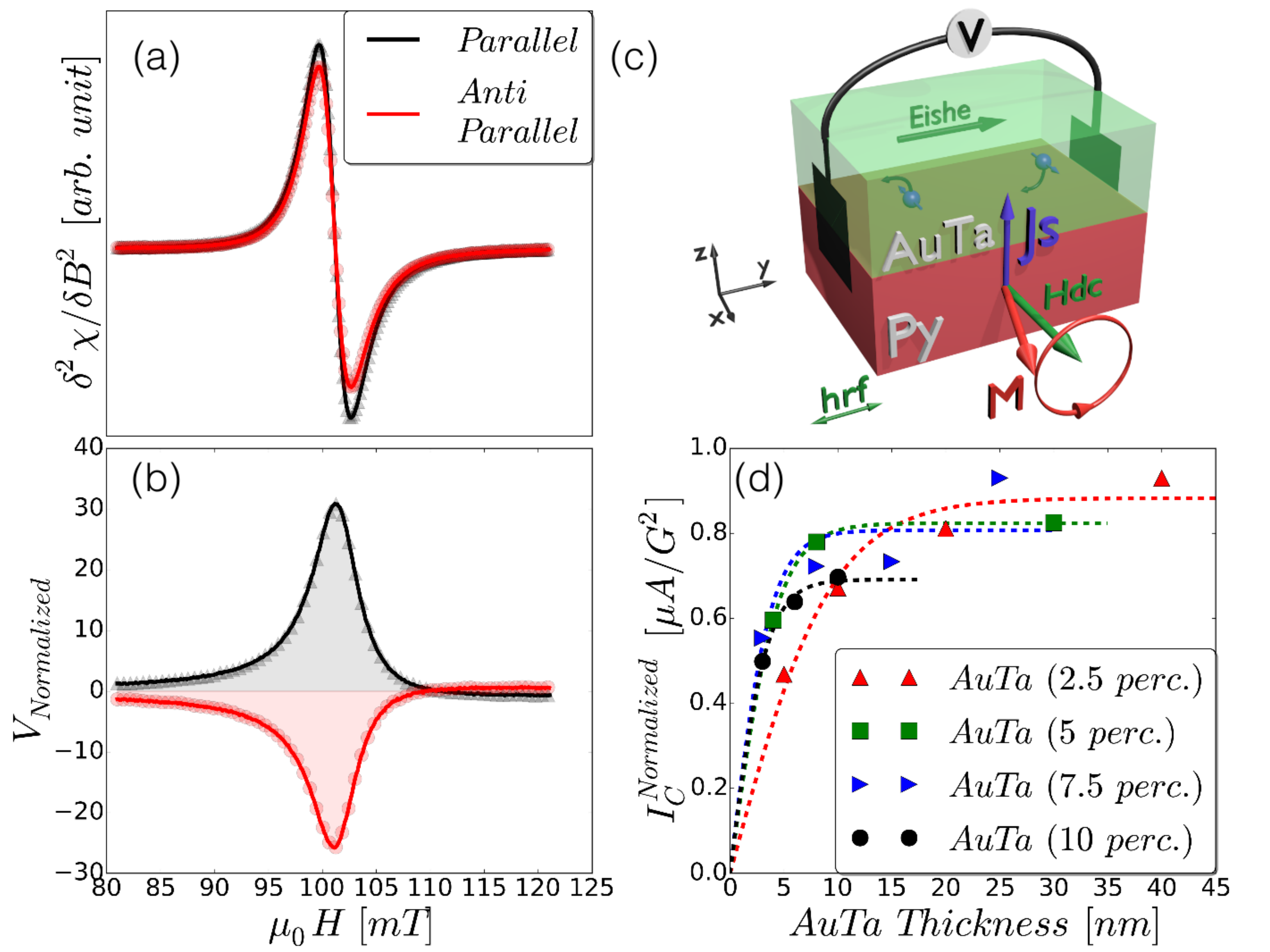}
		\caption{\textit{(a) and (b): Ferromagnetic resonance (FMR) spectra (a) and inverse spin Hall effect signal (b) of $AuTa2.5\%(40nm) \textbackslash NiFe(15nm)$ sample, measured in parallel and antiparallel positions (relative to the external magnetic field) in cavity, with $rf$ power of 50 mW. (c): Schematic representation of the experiment geometry. (d): AuTa thickness dependences of charge current production by ISHE for Ta concentrations of $2.5\%$, $5\%$, $7.5\%$, and $10\%$ and normalized in unit of rf field excitation, ${\mu}A/G^2$. Lines corresponds to fits using eq. (1). }}
\end{figure}

 The dependence of the charge current induced by conversion in AuTa, $I_{c}^{norm}$ (normalized in unit of rf field excitation, ${\mu}A/G^2$), is displayed in Fig. 3(d) for several samples. One can observe the expected first increase of $I_{c}^{norm}$ followed by its saturation as AuTa thickness t increases. The curves correspond to a fit to the expression \cite{MosendzPRB2010,RojasSanchez:2014ih}: 

\begin{equation}
	I_{c}^{norm}= \Theta_{SHE} \lambda_{sf} tanh (t/2\lambda_{sf}) J_s
\end{equation}

where $t$ and $J_s$ are the thickness of the Au alloys and the injected spin current density by spin pumping, respectively. $I_{c}^{norm}$ levels off at around $0.7 - 0.9 \,{\mu}A/G^2$ which is higher than what can we found for pure Pt ($0.6 \,{\mu}A/G^2$) \cite{RojasSanchez:2014ih}, W (-$0.45 \,{\mu}A/G^2$) or Ta (-$0.1 \,{\mu}A/G^2$) under similar experimental conditions and is a signature of the very high conversion rates for AuTa. The fit of the thickness dependence allows us to extract the spin diffusion lengths of the AuTa alloys shown in Fig. 4(a) up to $10\%$ of Ta content, see also \cite{Suppl}. As in the AuW alloy, the $\rho \times \lambda_{sf}$ product in AuTa is found to be constant in this resistivity range. For resistivities larger than $85\,\mu\Omega.cm$, the thickness dependence cannot yield reliable spin diffusion lengths $\lambda_{sf}$ as it becomes much smaller than the minimal SHE layer thickness of 4 nm. 
The SHA evaluated using eq. 1 is reported in Fig. 4(b) by taking into account the injected spin current density estimated by the FMR analysis, see \cite{Suppl}. The SHA increases almost linearly with the Ta content (and resistivity) and reaches a value as high as 50\% for concentrations in the 8-10 $\%$ range. As discussed in \cite{Suppl} and shown in Fig. 4(c), the SHA and spin diffusion lengths of our AuTa alloys allow a highly efficient spin to charge conversion. Note thant as the spin memroy loss is not yet included \cite{RojasSanchez:2014ih}, the SHA values are effective ones, lower bond of the intrinsic value.

\begin{figure}
	\centering{}
		\includegraphics[width=7.5cm]{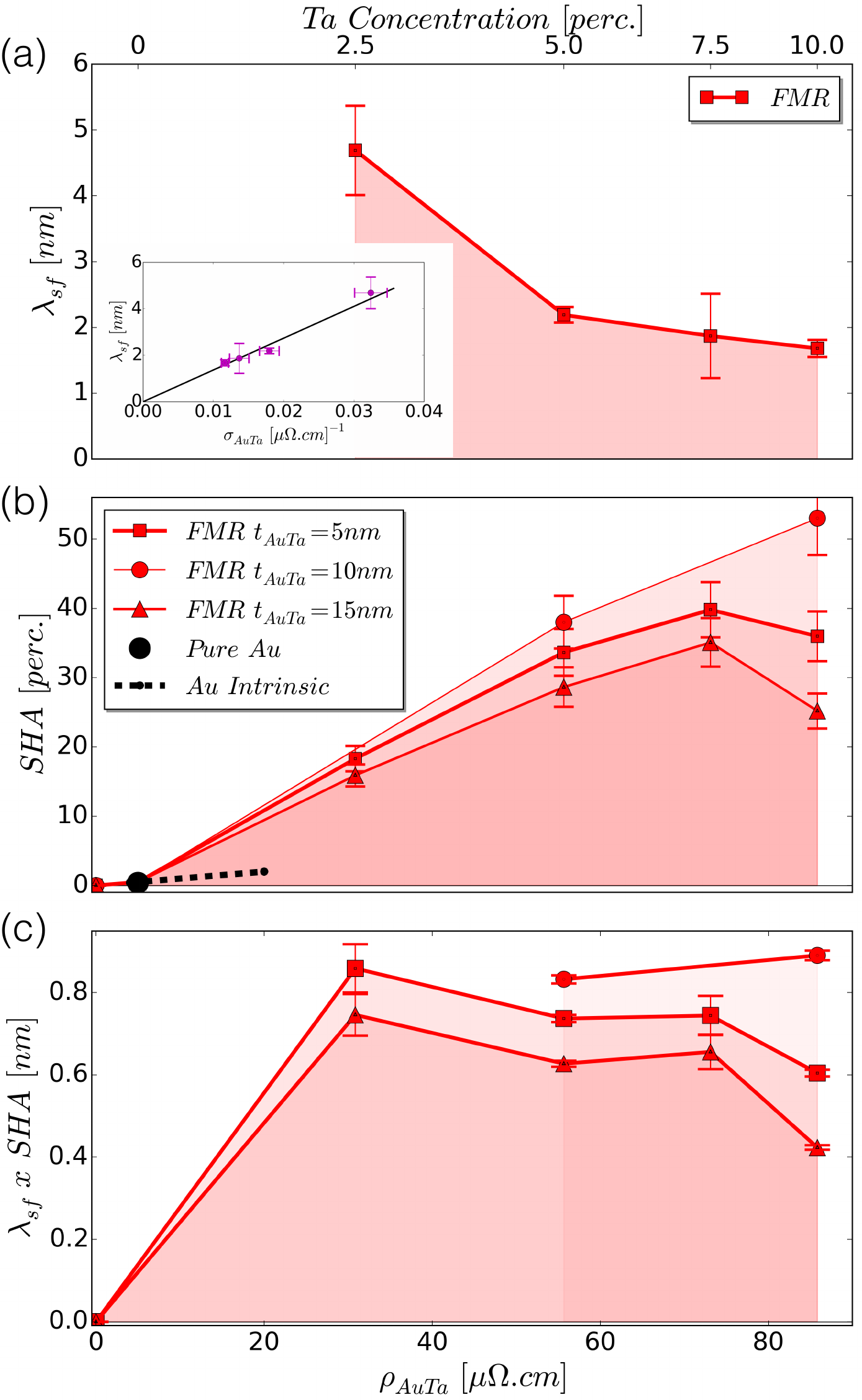}
		\caption{\textit{(a) Spin diffusion lengths extracted from the thickness dependence of the SP-FMR of Fig.3.(d) for the AuTa alloy. Inset evidences that $\rho \lambda_{sf}$ remains constant. (b) Spin Hall angle of AuTa alloys for different thickness of AuTa as a function of the Ta concentration (or AuTa resistivity). (c) Variation of the product of the spin Hall angle by the spin diffusion length characterizing the efficiency of the spin-charge conversion \cite{Suppl,Sanchez:kb}.}}
\end{figure}

The slope of the linear dependence of the SHA as a function of the longitudinal resistivity in AuTa is much steeper (by a factor of about 3) than the similar slope for AuW in Fig.2b or the slope expected for the intrinsic SHE at small concentration (dashed line in Fig.2b and 3b). This additional slope for Au doped with Ta can be attributed to side-jump scattering on Ta impurities, a mechanism which is also expected to provide a contribution to the SHA proportional to the resistivity \cite{Berger1970}. 
The existence of a large side-jump contribution to the SHE is confirmed by the calculation that we have worked out for alloys of $3\%$. We used an analysis based on combination of the resonant scattering Fert-Levy model~\cite{Fert:2011bl} and first principles calculations as used in case of BiCu~\cite{Niimi2012, Levy2013} or IrCu alloys~\cite{Xu2015}. Namely, the \textit{Ab-initio} calculations allow us to estimate the scattering phase shifts of s, p and d states involved in the model from the occupation numbers of these states according to Friedel's sum rule. For this purpose, the QUANTUM ESPRESSO \cite{QuantumEspresso} package with Perdew-Burke-Ernzerhof (PBE) pseudopotentials based on the generalised gradient approximation (GGA)~\cite{PBE} was used to calculate the number of electrons on the j=3/2 and j=5/2 states around W or Ta impurity in W$_1$Au$_{31}$ and Ta$_1$Au$_{31}$ supercells, respectively. By comparing with the number of electron on the Au atoms in a pure Au cell, all scattering phase shifts for the skew as well as their derivatives needed for the side-jump SHA contributions (see Ref.~\cite{Levy2013}) can be evaluated. For both alloys, the skew scattering is found to be negligible, the SHA being of the order of $10^{-4}$ for $~3\%$ content. These results are in-agreement with recent calculations based on alternative first-principles Kubo-Streda approach~\cite{Chadova}. This is also in line with the experiment where the intercept of the SHA slope at zero impurity content (or small resistivity) is found to be close to zero. On the contrary, for the side jump contribution, the SHA is found to be small for W but much larger for Ta. Starting from side-jump contribution calculated for a concentration of $3\%$ and after linear extrapolation to $10\%$ of Ta, we obtain a side-jump SHA equal to 0.33. Adding the contribution from intrinsic SHE estimated for Au at that resistivity, we finally obtain a SHA equal to 0.42 for $10\%$ Ta with a resistivity of $85 \mu\Omega.cm$, in good agreement with our experimental results. The saturation of the increase of SHA at concentration of Ta around $10\%$ anticipates the change of sign of the intrinsic term between Au and Ta. Note also that this concentration is close to the solubility limit for Ta in Au.

In summary, by studying Au based alloys, we have shown that impurities in a conducting host can be a suitable way to tune extensively the resistivity, the spin diffusion length and the spin Hall angle of materials. In AuTa alloys, thanks to a particularly large side-jump contribution, SHA as large as $50\%$ can be observed at Ta concentrations in the range of $10\%$. In comparison with the $\beta$-phase of pure W and Ta which also show large SHA, the smaller resistivity of the AuTa alloys is an advantage to obtain large spin currents and spin-torques with lower energy consumption in devices, as discussed in \cite{Suppl}. Our results open a new way in the quest of efficient spin-orbit materials for spintronics. 

We acknowledge the French ANR project SOspin and the Laboratoire d'excellence Lanef for financial support, as well as the renatech network, I. Pheng and L. Notin for sample fabrication.

%

\end{document}